\documentclass[aps,pre,twocolumn,showpacs,superscriptaddress]{revtex4-1}

\usepackage{amsmath,amssymb}
\usepackage{graphicx}
\usepackage{dcolumn}
\usepackage{bm}
\usepackage{color}
\usepackage[normalem]{ulem}

\begin{document}

\newcommand{\eq}[1]{Eq.~(\ref{#1})}  %example: \eq{eq:NFPE}

\title{Relaxation after a change in the interface growth dynamics}

\date{\today}

\author{T. A. de Assis}
\email{thiagoaa@ufba.br}
\address{Instituto de F\'{\i}sica, Universidade Federal da Bahia,
   Campus Universit\'{a}rio da Federa\c c\~ao,
   Rua Bar\~{a}o de Jeremoabo s/n,
40170-115, Salvador, BA, Brazil}

\author{F. D. A. Aar\~{a}o Reis}
\email{reis@if.uff.br}
\address{Instituto de F\'\i sica, Universidade Federal Fluminense, Avenida Litor\^anea s/n,
24210-340 Niter\'oi RJ, Brazil}

\begin{abstract}
The global effects of sudden changes in the interface growth dynamics are studied using models of
the Edwards-Wilkinson (EW) and Kardar-Parisi-Zhang (KPZ) classes during their growth regimes
in dimensions $d=1$ and $d=2$.
Scaling arguments and simulation results are combined to predict the relaxation of the difference in
the roughness of the perturbed and the unperturbed interfaces, $\Delta W^2 \sim s^c t^{-\gamma}$,
where $s$ is the time of the change and $t>s$ is the observation time after that event.
The previous analytical solution for the EW-EW changes is reviewed and numerically discussed in the
context of lattice models, with possible decays with $\gamma=3/2$ and $\gamma=1/2$.
Assuming the dominant contribution to $\Delta W^2$ to be predicted from a time shift in the final growth
dynamics, the scaling of KPZ-KPZ changes with $\gamma = 1-2\beta$ and $c=2\beta$ is predicted,
where $\beta$ is the growth exponent. Good agreement with simulation results in $d=1$ and
$d=2$ is observed.
A relation with the relaxation of a local autoresponse function in $d=1$ cannot be discarded, but
very different exponents are shown in $d=2$. We also consider changes between different dynamics,
with the KPZ-EW as a special case in which a faster growth, with dynamical exponent $z_i$, changes to a
slower one, with exponent $z$. A scaling approach predicts a crossover time $t_c\sim s^{z/z_i}\gg s$
and $\Delta W^2 \sim s^c F\left( t/t_c\right)$, with the decay exponent $\gamma=1/2$ of the EW class.
This rules out the simplified time shift hypothesis in $d=2$ dimensions.
These results help to understand the remarkable differences in EW smoothing of correlated and uncorrelated
surfaces, and the approach may be extended to sudden changes between other growth dynamics.
\end{abstract}

\maketitle

\section{Introduction}
\label{intro}

The study of kinetic roughening theory and related continuous and atomistic models
is motivated by the technological interest in thin films, multilayers and related nanostructures,
as well as the theoretical and experimental interest in fluctuating interface problems
\cite{barabasi,krug,etb,ohring}.
Some important classes of interface growth are those connected to the Edwards-Wilkinson (EW) equation
\cite{ew}, in which linear interface tension is the dominant relaxation term, and to the
Kardar-Parisi-Zhang (KPZ) equation \cite{kpz}, which includes a nonlinear effect of the local slope.
Recent advances in the solution of the KPZ equation
\cite{sasamoto,calabrese,imamura,healy2012,tiago2012,healy2013,tiago2013}
and some experimental realizations \cite{takeuchi,yunker,renan} renewed the interest in those problems.

A relatively small number of works considered thin film and interface growth problems with
time-dependent conditions \cite{shapir,foster,pradas,anomcompet,tempvar},
although there is a large number of experimental problems with that feature, ranging from
fluid imbibition in porous media \cite{pradas,dube,alava} to thin film electrodeposition \cite{foster,PB}.
Many of those models and experiments show anomalous scaling of the surface roughness \cite{ramasco}.
A more recent application is the deposition of compositionally graded films, in which the flux of
different species vary in time \cite{zhouASS,vyas,zhangASS,fuentes}.
This technique may improve film adhesion and reduce internal stress, among other benefits.

Another possibility is a sudden change in the dynamics during the interface growth.
For instance, this is the case of a change from sputtering to annealing in a cycle of surface
cleaning \cite{ohring}. Moreover, any change from
a surface cleaning process to thin film deposition on that surface may be viewed as a potential
application. Since erosion or dissolution are frequently present in those processes, we recall
that KPZ scaling was already observed in several etching and dissolution models
\cite{mello,passalacqua,mack,flavio2013}.
On the other hand, KPZ scaling were also observed in $LiCoO_x$ films after high temperature
annealing, with initial deposition by sputtering \cite{kleinke1999}.
Thus, the apparently simple situation of a sudden change in the growth dynamics may have a
variety of applications that involve KPZ scaling.

The problem of changes in the EW equation in dimension $d=1$ was studied in Ref.
\protect\cite{chou} and further extended to other linear growth equations
in all $d$ \cite{chouJSTAT}. Those works showed power-law relaxation of
the difference $\Delta W^2$ of the square roughness of the perturbed and the unperturbed system,
which measures the global response to the perturbation  \cite{chou,chouJSTAT}.
This feature may be important for experimental works in
which there is any sudden change in conditions such as temperature, pressure and composition,
since a delay in the response to a change may affect the desired film properties.
Indeed, the slow relaxation referred above was observed when both the initial and the final EW dynamics
were in the growth regime, which corresponds to typical experimental conditions,
in contrast to the exponential relaxation observed in steady state (very long time) properties.
On the other hand, recent works studied autoresponse functions in KPZ models \cite{henkel,odorcondmat},
which measure the average local response to a perturbation and show particular aging properties.
This is an additional reason to search for a deeper understanding of the relaxation of global
quantities, in particular when KPZ growth is involved.

The aim of this work is to study the effect of sudden changes in the EW and KPZ dynamics,
including changes between these different growth classes. The scaling of the global quantity $\Delta W^2$
is analyzed, with support from simulation results for a variety of lattice models in $d=1$ and $d=2$.
A previous analytical solution for the EW-EW changes is reviewed and provides the background for
a simple scaling approach to the KPZ-KPZ changes, in which the dominant contribution to $\Delta W^2$
is predicted from a time shift in the final growth dynamics.
The striking difference from local autoresponse functions is clearly shown for KPZ in $d=2$.
For KPZ-EW changes, crossover times significantly exceed the growth time with the initial dynamics,
which is an expected general trend when a faster dynamics is changed to a slower one (corresponding to an
increase in the dynamic exponent). Moreover, remarkable differences in EW smoothing of correlated
and uncorrelated surfaces are discussed.

This paper is organized as follows. In Sec. \ref{models}, we present basic definitions,
the interface equations and the lattice models considered in this work.
In Sec. \ref{ewew}, we briefly review previous analytical results on changes between initial
and final EW growth, define a suitable scaling function and propose an approach to
explain the simplest type of relaxation. In Sec. \ref{kpzkpz}, that approach is extended to
changes from initial and final KPZ growth.
In Sec. \ref{kpzew}, we introduce a scaling approach
for the KPZ-EW changes, which are confirmed by numerical results in $d=1$ and $d=2$.
In Sec. \ref{smoothing}, we discuss the smoothing of very rough surfaces by EW dynamics.
Sec. \ref{conclusion} summarizes our results and conclusions.

\section{Basic definitions, interface equations, and lattice models}
\label{models}

The simplest quantitative characteristic of an interface is its roughness (or interface width).
It is usually defined as the rms fluctuation of the height $h$ as
\begin{equation}
W(L,t)\equiv
{\left[ { \left<  \overline{{\left( h - \overline{h}\right) }^2}  \right> } \right] }^{1/2} ,
\label{defW}
\end{equation}
where $L$ is the lateral size and $t$ is the growth time.
The overbars in Eq. (\ref{defW}) represent spatial averages and the angular brackets represent configurational
averages. The roughness can be calculated from the structure factor
\begin{equation}
S(\vec{k},t)\equiv
{ \left<  \tilde{h}\left( \vec{k},t\right) \tilde{h}\left( -\vec{k},t\right) \right> }
\label{defS}
\end{equation}
as
\begin{equation}
W^2(L,t)\equiv \sum_{\vec{k}}{S(\vec{k},t)} ,
\label{WS}
\end{equation}
where $\tilde{h}$ is the Fourier transform of $h$ given by
$\tilde{h}\left( \vec{k},t\right)=\sum_{\vec{r}}{{h\left( \vec{r},t\right)}e^{i\vec{k}\cdot \vec{r}}}$
($r$ denotes the position in $d$ dimensions and $\vec{k}$ is the wave vector).

In interface growth processes with normal scaling (in opposition to anomalous scaling \cite{ramasco}),
the roughness follows the Family-Vicsek scaling relation \cite{fv}
\begin{equation}
W \approx L^{\alpha} f\left( \frac{t}{t_\times}\right) ,
\label{fv}
\end{equation}
where $\alpha$ is the roughness exponent, $f$
is a scaling function such that $f\to const$ in the regime of roughness saturation
($t\to\infty$) and $t_\times$ is the characteristic time of crossover to
saturation. $t_\times$ scales with the system size as
\begin{equation}
t_\times \sim L^z ,
\label{defz}
\end{equation}
where $z$ is the dynamic exponent.
For $t\ll t_\times$ (but after a possible transient), the roughness scales as
\begin{equation}
W\sim t^\beta ,
\label{defbeta}
\end{equation}
where $\beta=\alpha/z$ is the growth exponent. In this
growth regime, the structure factor scales as
\begin{equation}
S(k,t) \sim k^{-\left( 2\alpha+d\right)} g\left( k^zt\right) ,
\label{scalingS}
\end{equation}
where $g$ is a scaling function.

In this work, our interest is to study the interface evolution in the growth regime, with
negligible finite-size effects. The roughness of the interface with a sudden change of dynamics at time $s$
is referred as $W_c(t,s)$ and the roughness of the interface grown with
the final dynamics since $t=0$ is referred as $W_u(t,s)$.
The exact result for EW-EW changes in $d=1$ \cite{chou,chouJSTAT} suggests to define
a reduced time as
\begin{equation}
\tau\equiv t/s-1 .
\label{deftau}
\end{equation}
A general scaling form for the roughness difference
\begin{equation}
\Delta W^2(t,s)\equiv |W_c^2 -W_u^2| .
\label{defdW2}
\end{equation}
between the changed and unchanged systems is
\begin{equation}
\Delta W^2 \sim s^c \tau^{-\gamma} .
\label{defcgamma}
\end{equation}
The KPZ equation is
\begin{equation}
{{\partial h}\over{\partial t}} = \nu{\nabla}^2 h + {\lambda\over 2}
{\left( \nabla h\right) }^2 + \eta (\vec{r},t) ,
\label{kpz}
\end{equation}
where $h$ is the interface height at the position $\vec{r}$ in a
$d$-dimensional substrate at time $t$, $\nu$ represents the surface tension,
$\lambda$ represents the excess velocity and $\eta$ is a Gaussian
noise \cite{barabasi,kpz} with zero mean and co-variance $\langle
\eta\left(\vec{r},t\right) \eta (\vec{r'},t')\rangle = D\delta^d
(\vec{r}-\vec{r'} ) \delta\left( t-t'\right)$.
The EW equation \cite{ew} corresponds to the KPZ equation with $\lambda =0$, while
uncorrelated Growth (UG) is obtained for $\nu =0$ and $\lambda =0$.

The exact solution of the EW equation gives $z=2$ and $\alpha = (2-d)/2$ for $d\leq 2$
($\alpha =0$ in $d=2$ corresponding to logarithmic scaling) \cite{ew}.
In $d=1$, the KPZ equation has $z=3/2$ and $\alpha=1/2$  \cite{kpz};
in $d=2$, the best current numerical results give $z\approx 1.61$ and $\alpha\approx 0.39$
\cite{marinari,kpz2d,Kelling}. UG has $\beta=1/2$ and no roughness saturation, so that
$\alpha$ and $z$ are not defined.

Many lattice models share the same scaling exponents with EW or KPZ equations and
are said to belong to the EW or to the KPZ class. These models are expected to be represented by those
equations in the continuous limit (very large sizes, very long times), possibly with additional
higher order spatial derivatives that are irrelevant under renormalization \cite{barabasi}.

In all models studied here, the growth begins with a flat substrate at $t=0$.
Lattice sizes are $L=2^{14}$ in $d=1$ and $L=2^{10}$ in $d=2$. One time unit corresponds
to $L^d$ deposition trials (deposition of one layer of particles in solid on solid models).
Maximal growth times are chosen well below the saturation regime, except if explicitly indicated.
Changes take place at time $s$, with $s$ varying from $10$ to $10^{3}$.

The lattice model in the EW class studied here is the Family model \cite{family}.
At each step of this model, a column of the deposit is randomly chosen and the minimum height
is searched up to a distance $N$ from that column. If no column in that neighborhood has a
height smaller than that of the column of incidence, a new particle sticks at the top of this one.
Otherwise, it sticks at the top of the column with the smallest height in that neighborhood.
If two or more columns have the same minimum height, the sticking position is the one closest
to the incidence column and, in the case of a new draw, one of the smallest and closest columns
is randomly chosen.
The increase of $N$ corresponds to an increased interface tension compared to the noise
intensity, i. e. an increase of the ratio $\nu /D$ in the corresponding EW equation.
Hereafter, the Family model with searching distance $N$ will be referred as F$N$ model.

The KPZ models considered here are the restricted solid-on-solid (RSOS) model \cite{kk}
and the etching model of Mello et al \cite{mello}. The latter is particularly interesting
due to the large number of applications of etching processes (by aggressive solutions, sputtering, etc).

In the $RSOS$ model \cite{kk}, the incident particle may stick at the
top of the column of incidence if the differences of heights between
the incidence column and each of the neighboring columns
do not exceed ${\Delta h}_{MAX} = 1$. Otherwise, the
aggregation attempt is rejected.

The model for etching of a crystalline solid of Mello et al \cite{mello} is
simulated here in its deposition version, hereafter called ETCH model.
At each deposition attempt, the height of the
column of incidence is increased by one unit ($h(i)\rightarrow h_0+1$) and any neighboring
column whose height is smaller than $h_0$ grows until its height becomes $h_0$
(in the true etching version of this model, the columns' heights decrease by the same
quantities above).

Finally, the UG is simulated with aggregation at the top of the column of incidence, independently of
the neighboring heights.

\section{Changes EW-EW}
\label{ewew}

One-dimensional EW growth with a sudden changes in the interface tension was studied by Chou et al \cite{chou}.
Subsequently, Chou and Pleimling extended that approach to changes in interface tension and noise
amplitude in any dimension \cite{chouJSTAT}. Considering initial dynamics with parameters $(\nu_i,D_i)$
and the final one with $(\nu_f,D_f)$, the difference in the square roughness can be written as \cite{chou,chouJSTAT}
\begin{equation}
\Delta W^2(t,s)\equiv |W_c^2 -W_u^2|\sim \sum_{\vec{k}}\frac{e^{-2\nu_f k^2\left( t-s\right)}}{k^2}
\Delta G_{EW}\left( k^2 s\right)  ,
\label{dW2EWEW}
\end{equation}
with
\begin{equation}
\Delta G_{EW}\left( k^2 s\right) = \left| D_i\frac{1-e^{-2\nu_i k^2 s}}{2\nu_i} -D_f\frac{1-e^{-2\nu_f k^2 s}}{2\nu_f} \right| .
\label{dGEWEW}
\end{equation}
With this form, $\Delta G_{EW}\left( 0\right)=0$.
If the initial and the final dynamics are in the growth regime at times $s$ and $t$, respectively, then
$k^2s,k^2t\ll 1$. The lowest nonzero order in the expansion of $\Delta G_{EW}\left( x\right)$ depends on
the type of change: if only $D$ is changed ($\nu_i=\nu_f$), then it is first order in $x\equiv k^2s$;
if only $\nu$ changes ($D_i=D_f$), the leading order is the second one.
In any case, $\Delta W^2(t)$ can be written in terms of the scaling variable $u\equiv k^2s\tau$
[see Eq. \ref{deftau}], which gives $\gamma = d/2$ (change in $D$) and $\gamma = d/2+1$ (change in $\nu$),
with $c=1-d/2$ in both cases [see Eq. (\ref{defcgamma})].

In $d=1$, changing only interface tension leads to $c=1/2$ and $\gamma =3/2$.
Numerical results of Chou et al \cite{chou} gave exponent $\gamma\sim 3/2$ typically for
$\nu_i/\nu_f< 10$ and $\nu_f s>0.1$, which corresponds to an initial interface tension not much larger
than the final one. A limiting case in lattice models is the UG-F1 change, where $\nu_i =0$.
Indeed, Fig. 1a shows $\Delta W^2(t) s^{-c}$ versus $\tau$, using the
value $c=0.42$ that provides the best data collapsed for three different value of $s$. It confirms
the predicted slope $-3/2$, with some deviations only for large $\tau$.

\begin{figure}[!h]
\includegraphics [width=10cm] {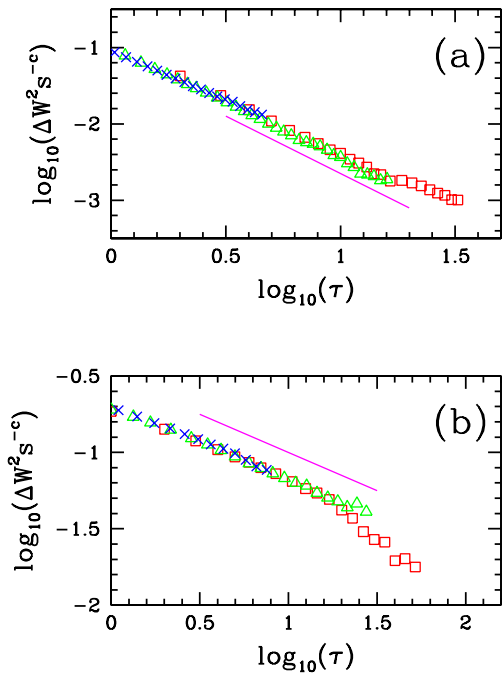}
\caption{(Color online) Scaling of the roughness difference in the changes: (a) UG-F1 with
$s=10$ (red squares), $s=100$ (green triangles), and $s=1000$ (blue crosses), using $c=0.42$;
(b) F50-F1 with times $s=10$ (red squares), $s=30$ (green triangles), and $s=100$ (blue crosses),
using $c=0.48$.
The solid lines in (a) and (b) have slope $-3/2$ and $-1/2$, respectively.}
\label{fig1}
\end{figure}
On the other hand, Chou et al \cite{chou} obtained $\gamma\approx 1/2$ in a wide region with
$\nu_i/\nu_f>{10}^4$ and $\nu_f s<1$, with fixed $D$.
This condition typically corresponds to a very large initial interface
tension, which produces a smooth surface that rapidly brings the interface to a steady state
of very low roughness. In lattice models, this is illustrated by the change F50-F1, which is shown in
Fig. 1b. The best data collapse is obtained with $c=0.48$ and the slope of the plot is close to $-1/2$
(some deviations appearring only for $s=10$ at long times, in which the accuracy of $\Delta W^2$ is low).

The form of  $\Delta G_{EW}$ in Eq. (\ref{dGEWEW}) helps to understand this result. The
first contribution to $\Delta G_{EW}$ vanishes in the saturation regime of the F50 model and, consequently,
$\Delta G_{EW}(x)$ is of first order for small $x$. It gives $c=1/2$ and $\gamma = 1/2$,
in agreement with the numerical estimates. This is similar to the case of changing the
noise amplitude, although in F50-F1 we understand that only a change in interface tension is present.

An equivalent reasoning that leads to a first order dominant term in $\Delta G_{EW}$ and its
corresponding exponents
is to assume that $\Delta W^2(t)$ is dominated by a difference of roughnesses of the final dynamics
with starting times $0$ and $s$. This gives

\begin{equation}
\Delta W^2(t,s)\approx At^{2\beta}-A{\left( t-s\right)}^{2\beta} \sim st^{2\beta-1} .
\label{timeshiftEW}
\end{equation}

Since $\beta=\left( 2-d\right)/4$ for EW growth \cite{ew}, we obtain $c=1-d/2$ and $\gamma=d/2$.
This is certainly a good approximation when the roughness at $t=s$ is very small, which is the
case of the F50 model. However, it also applies when the initial roughness is not small.
In this case, the initial and final dynamics have the same dynamic exponent $z$, thus the initial correlations,
created in a time $s$, are changed by the final dynamics in a time of the same order. For this reason,
$\Delta W^2(t,s)$ is approximately related to a difference of starting times of order $s$.
In this context, the case $D_i=D_f$, $\nu_i\neq \nu_f$ can be understood as a particular case in which this
first order correction vanishes and a more rapid decay is observed.

\section{Changes KPZ-KPZ}
\label{kpzkpz}

The relaxation of $\Delta W^2\left( t,s\right)$ in the integrated KPZ equation in $d=1$ was
formerly studied by Chou and Pleimling \cite{chouJSTAT}, who obtained the scaling
relation (\ref{defcgamma}) with $\gamma = 1/3$ and $c=2/3$.
Now, we will show that, interestingly, those results are predicted by the
same scaling argument that leads to Eq. (\ref{timeshiftEW}), now with KPZ exponents.
Moreover, our arguments can be extended to $d>1$, as follows.

For $t\gg s$, $\Delta W^2\left( t,s\right)$ for KPZ-KPZ changes is written as a difference of roughnesses
similarly to Eq. (\ref{timeshiftEW}). This gives the scaling form (\ref{defcgamma}) with
\begin{equation}
\gamma = 1-2\beta , c=2\beta ,
\label{gammabetaKPZKPZ}
\end{equation}
with exponent $\beta$ defined in Eq. (\ref{defbeta}).

Fig. 2 shows $\Delta W^2\left( t,s\right) s^{-0.66}$ versus $\tau$ for the changes RSOS-ETCH and ETCH-RSOS
in $d=1$, with three different times $s$. The good data collapse and the long time slope
near $-1/3$ confirms the above assumptions.
It also agrees with the numerical results of Ref. \protect\cite{chouJSTAT}.

\begin{figure}[!h]
\includegraphics [width=8.0cm] {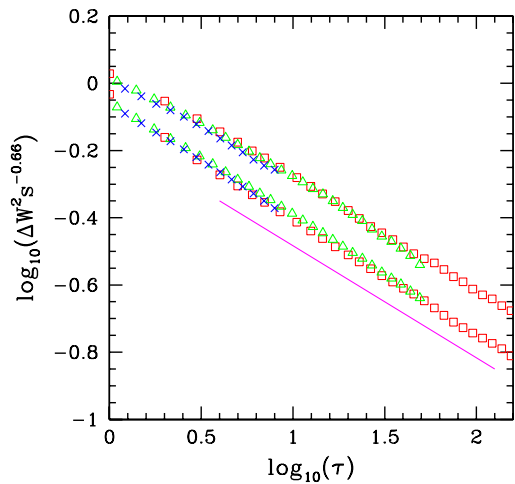}
\caption{(Color online) Scaling of the roughness difference in the changes RSOS-ETCH (upper data)
and ETCH-RSOS (lower data) in $d=1$, with $s=10$ (red squares), $s=100$ (green triangles), and
$s=1000$ (blue crosses). The solid line has slope $-1/3$.
}
\label{fig2}
\end{figure}

\begin{figure}[!h]
\includegraphics [width=10cm] {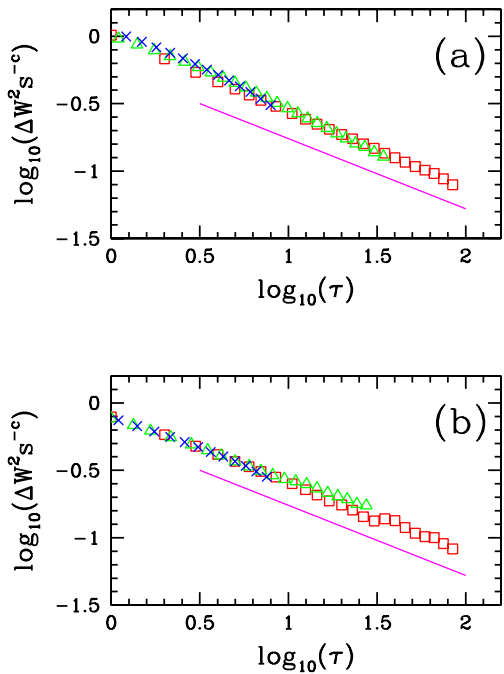}
\caption{(Color online) Scaling of the roughness difference in $d=2$:
(a) ETCH-RSOS with $c=0.42$ and changes at $s=10$ (red squares), $s=100$ (green triangles), and
$s=1000$ (blue crosses); (b) RSOS-ETCH with $c=0.45$ and changes at $s=10$ (red squares),
$s=30$ (green triangles), and $s=100$ (blue crosses). The solid lines in (a) and (b) have slopes $-0.52$.}
\label{fig3}
\end{figure}
The scaling of the difference of the roughness $W$ (not squared) can be obtained
from Eq. (\ref{defcgamma}) by noting that
$\Delta W^2 = |W_c^2 -W_u^2| = |W_c -W_u|\left( W_c +W_u\right)$, with $W_c$ and $W_u$ scaling
as Eq. (\ref{defbeta}) for large $t=s\tau$. This gives
\begin{equation}
\Delta W (t,s)\equiv | W_c(t,s)-W_u(t) |\approx s^{\beta} \tau^{\beta-1} ,
\label{dWts}
\end{equation}
where exponents in Eq. (\ref{gammabetaKPZKPZ}) were used.
In $d=1$, we obtain $\Delta W \sim s^{1/3}\tau^{-2/3}$.

In Ref. \protect\cite{henkel}, an autoresponse function was defined from the
average differences in local heights between two interfaces, A and B, the former growing
with a site-dependent rate up to time $s$ and, after that, with uniform rate,
and the latter growing with site-independent rates from $t=0$. That function is given by:

\begin{equation}
\chi \left(s,t\right) =
\overline{\left\langle \frac{h^{(A)}_i\left( t,s\right)-h^{(B)}_i\left( t\right)}{\epsilon_i}\right\rangle},
\end{equation}
where $i$ refers to lattice columns and $\epsilon_i$ is proportional to the (small) fluctuation
in the growth rate at column $i$.
Thus, $\chi \left(s,t\right)$ measures the average local response to a small perturbation.
Surprisingly, $\chi \left(s,t\right)$ has the same scaling as
$\Delta W (t,s)$ [Eq. (\ref{dWts})], which is a difference in a global quantity subject to
a uniform change in growth parameters.

In $d=2$, we also analyzed the changes ETCH-RSOS and RSOS-ETCH  by plotting $\Delta W^2(t) s^{-c}$
versus $\tau$ for three times $s$ and searching for the values of $c$ that provide the best
data collapses. The corresponding scaling plots are shown in Figs. 3a and 3b, respectively with
$c=0.42$ and $c=0.45$, in good agreement with Eq. (\ref{timeshiftEW}) that
predicts $c=0.48$ from $\beta=0.24$ \cite{Kelling}. The predicted slope $\gamma = 0.52$
is also shown in Figs. 3a and 3b, confirming the scaling of Eq. (\ref{timeshiftEW}).

The autoresponse function $\chi \left(s,t\right)$ was recently studied in a KPZ model in $d=2$
by \'Odor et al \cite{odorcondmat}, who obtained $\chi\approx s^{0.3}f_r\left( \tau\right)$ and
$f_r\left( x\right)\sim x^{-1.25}$ for large $x$. This scaling is completely different from
the scaling of $\Delta W^2\left( t,s\right)$ and of $\Delta W\left( t,s\right)$.
This shows the striking differences among the local and global responses in $d=2$.

The scaling of $\Delta W^2\left( t,s\right)$ for the EW-EW changes was based on
Eqs. (\ref{dW2EWEW}) and (\ref{dGEWEW}). Thus, the rest of this section is devoted to investigate
the consequences of assuming similar relations for KPZ-KPZ changes.
However, we stress that the following reasoning is based on speculations on KPZ scaling that
cannot be justified by current analytical works on the subject.

Our first step is to replace the scaling of the structure factor in Eq. (\ref{dW2EWEW})
by that of KPZ, with corresponding exponents $z$ and $\alpha$.
Secondly, a function $\Delta G$ is also used to represent the effect of
the change of dynamics on the mode $k$. These assumptions give

\begin{equation}
\Delta W^2(t,s)\approx \sum_{\vec{k}} g[k^z\left( t-s\right)]{k^{-(2\alpha+d)}}
\Delta G_{KPZ}\left( k^z s\right)  .
\label{dW2KPZKPZ}
\end{equation}
for KPZ-KPZ changes. This is equivalent to assume that the scaling of $\Delta W^2$ is not dominated
by coupling of different modes (while in EW scaling there is no mode coupling at all).

Now we also assume that  the leading nonzero order of $\Delta G_{KPZ}(x)$ is the first one,
i. e. $\Delta G_{KPZ}(x)\sim x$ for small $x$. Eq. (\ref{dW2KPZKPZ})
can be rewritten in terms of the variable $k^z s\tau$ and gives the scaling of Eq. (\ref{timeshiftEW})
with the exponents in Eq. (\ref{gammabetaKPZKPZ}). We also note that any other assumption for
the leading order of $\Delta G_{KPZ}(x)$ would provide different exponents.

Thus, the assumption of a very simple scaling form for $\Delta W^2\left( t,s\right)$
[Eq. (\ref{dW2KPZKPZ}) with $\Delta G_{KPZ}(x)\sim x$] leads to the correct exponents for the KPZ-KPZ decay.
This is not an actual calculation for KPZ, but  may motivate the development of rigorous approaches for the subject.

\section{Changes to a different growth class}
\label{kpzew}

Now we consider a problem not addressed in previous works, which mainly corresponds to turning in or
turning off the nonlinearity in EW-KPZ or KPZ-EW changes, respectively.

Since KPZ correlations are spread faster than EW correlations
(e. g. $z_{KPZ}=3/2$ and $z_{EW}=2$ in $d=1$ \cite{ew,kpz}),
the correlation length of KPZ growth at $t=s\ll 1$ is much larger than the
correlation length of EW growth. Thus, in a change EW-KPZ, the time $t-s$ necessary for the
KPZ growth to supress the initial EW correlations is smaller than $s$. This leads to a crossover
scaling similar to the KPZ-KPZ changes discussed in Sec. \ref{kpzkpz}.

On the other hand, if the initial dynamics is KPZ,
the time necessary for the correlations at $t=s$ to be replaced by EW correlations will be
significantly larger than $s$. A suitable scaling approach has to be developed, along the same
lines of related approaches for EW-KPZ crossover of roughness scaling \cite{ggg,nt,forrest,tiago1}.
Hereafter we refer to the scaling exponents of the initial dynamics with subindex $i$
($z_i$, $\alpha_i$, $\beta_i$) and to those of the final dynamics with no subindex, so that
the approach may be easily extended to other growth classes.

At time $s$, the correlation length of the KPZ interface is $l_i\sim s^{1/z_i}$ and the
square roughness is
\begin{equation}
{W_i}^2 \left( s\right)\sim s^{2\beta_i} .
\label{wi}
\end{equation}
For $s\gg 1$, the final dynamics is so that ${W_u}^2\left( s\right)\sim s^{2\beta}\ll {W_i}^2 \left( s\right)$,
given that $\beta_i>\beta$, which is always the case in KPZ-EW changes. Thus, the change
produces a significant decrease in the roughness.

After the change, the correlation length of EW grows as $l_f\sim {\left( t-s\right)}^{1/z}$.
A crossover from the initial to the final dynamics is expected as $l_i\sim l_f$,
which means that initial KPZ correlations were replaced by EW correlations. The crossover time
$t_c$ scales as
\begin{equation}
t_c \sim s^{z/z_i}
\label{tcKPZEW}
\end{equation}
and a properly defined crossover variable is
\begin{equation}
y\equiv {\left( t-s\right)}/t_c .
\label{defy}
\end{equation}
This variable plays the role of the scaled time $\tau$ of Eq. \ref{deftau}.
We expect the difference in the square roughness to scale as
\begin{equation}
\Delta W^2 \approx s^cF\left( y\right) ,
\label{defy}
\end{equation}
where $F$ is a scaling function.

For $t\approx s$, $y\ll 1$ and using Eq. (\ref{wi}) we have
$\Delta W^2\left( t,s\right)\approx {W_i}^2\left( t,s\right)-{W_u}^2\left( t,s\right)
\approx s^{2\beta_i}$.
Thus,
\begin{equation}
c=2\beta_i
\label{cKPZEW}
\end{equation}
and $F(y)\to const$ in that limit.

At long times ($t\gg t_c$), $\Delta W^2(t)$ is expected to decay according to the final EW scaling.
This gives
\begin{equation}
F(y)\sim y^{-\gamma}, \gamma=d/2
\label{Fygg1}
\end{equation}
for $y\gg 1$, as in the case of first order dominant term in $\Delta G_{EW}$ [Eq. (\ref{dGEWEW})].

\begin{figure}[!h]
\includegraphics [width=8.0cm] {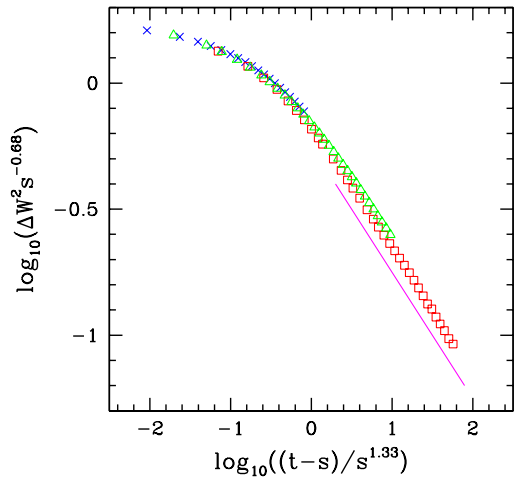}
\caption{(Color online) Scaling of the roughness difference in the change ETCH-F1 in $d=1$,
with $s=10$ (red squares), $s=100$ (green triangles), and
$s=1000$ (blue crosses). The solid line has slope $-1/2$.}
\label{fig4}
\end{figure}

In $d=1$, using EW and KPZ exponents \cite{ew,kpz},
this scaling approach gives $t_c\sim s^{4/3}$, $c=2/3$, and $\gamma =1/2$.
It is confirmed in Fig. 4, in which  $\Delta W^2(t) s^{-0.68}$ is plotted as a function of
$\left( t-s\right) /s^{1.33}$ for the change ETCH-F1 at three different times $s$.
The value of $c$ in Fig. 4 was chosen to provide the best data collapse.
Also note the trend of the scaling function to be flat
as $\left( t-s\right) /s^{1.33}\ll 1$, as predicted above.
The change RSOS-F1 is not analyzed here because the roughness of RSOS for short times is
smaller than that of F1, which invalidates the assumptions of the theoretical approach.

In $d=2$, $z=2$ \cite{ew}, $z_i\approx 1.61$, and $\beta\approx 0.24$ \cite{kpz2d,Kelling}
give $t_c\sim s^{1.24}$, $c\approx 0.48$, and $\gamma =1/2$.
Fig. 5 shows  $\Delta W^2(t) s^{-0.51}$ as a function of
$\left( t-s\right) /s^{1.24}$ for the change ETCH-F1 at three different times $s$,
again with the value $c=0.51$ chosen to provide the best data collapse.
The trend of the scaling function to be flat as $\left( t-s\right) /s^{1.24}\ll 1$ is
also noticeable in Fig. 5.
These results are in good agreement with our scaling approach, except for deviations
in the estimates of exponent $c$ used to get data collapse, which is already expected from
the experience with (exactly solved) EW-EW changes (Sec. \ref{ewew}).

\begin{figure}[!h]
\includegraphics [width=8.0cm] {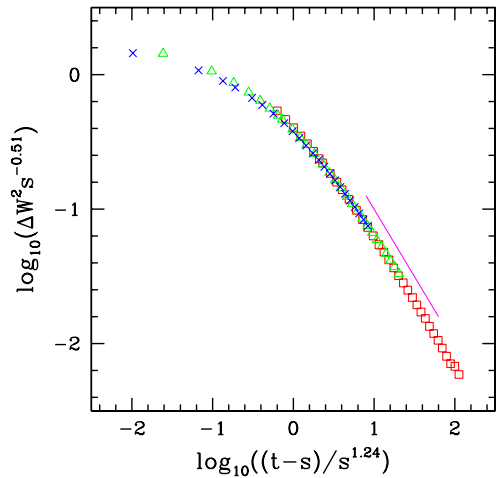}
\caption{(Color online) Scaling of the roughness difference in the change ETCH-F1 in $d=2$,
with $s=10$ (red squares), $s=20$ (green triangles), and
$s=40$ (blue crosses). The solid line has slope $-1$.}
\label{fig5}
\end{figure}

The hypothesis that $\Delta W^2(t)$ is dominated by a difference of roughnesses increasing
from zero at times $0$ and $t_c$ [equivalent to Eq. (\ref{timeshiftEW})] is not obvious in this case,
since there may be a significant roughness reduction in the KPZ-EW transition.
That hypothesis is
\begin{equation}
\Delta W^2\approx At^{2\beta}-A{\left( t-t_c\right)}^{2\beta} \sim t_c t^{1-2\beta} ,
\label{timeshiftKPZEW}
\end{equation}
which gives the same decay in Eq. (\ref{Fygg1}). However, it gives $c=2\alpha/z_i=\left( 2-d\right) /z_i$
in Eq. (\ref{defy}), in contrast to Eq. (\ref{cKPZEW}).
In $d=1$, this time shift hypothesis gives the correct value of $c$ because the roughness exponents of
EW and KPZ are the same. However, in $d=2$, it gives $c=0$, in striking disagreement with the
previous approach and the numerical data in Fig. 5.

For the time shift hypothesis to be valid, it is necessary that $\Delta W^2\ll {W_u}^2$,
i.e. the roughness difference has to be smaller than the roughness of the unperturbed growth.
For $t\geq t_c$, Eqs. (\ref{defy}), (\ref{cKPZEW}), and (\ref{Fygg1}) give
$\Delta W^2\sim s^{2\beta_i} {\left( t_c/t\right)}^{d/2}$, while ${W_u}^2\sim t^{1/2}$
in $d=1$ and ${W_u}^2\sim \log{t}$ in $d=2$ \cite{ew}. In $d=1$, $\Delta W^2\ll {W_u}^2$ require $t\gg t_c$;
the numerical results in Fig. 4 for $t/t_c>10$ are sufficient to satisfy this condition.
However, in $d=2$, that relation requires $t\gg t_c s^{0.48}\gg t_c$ (for $s\gg 1$ and
excluding a logarithmic correction in $s$); this condition is very far from the limits of the data in Fig. 5
(instead, the data in Fig. 5 typically has ${W_u}^2 < \Delta W^2$).
This explains the failure of the time shift hypothesis for the KPZ-EW change in $d=2$.

\section{EW smoothing of initially rough surfaces}
\label{smoothing}

Due to its logarithmic growth in time, the roughness of an EW interface in $d=2$ is very small
at all times representative of a thin film growth. This is the case of the maximal times
$t={10}^3$ considered in the data shown of Fig. 5. Moreover, until $t\sim {10}^7$
(ten millions of layers), the roughness of the Family model is smaller than two lattice units.
For this reason, it is interesting to compare
the effects of an EW smoothing of initially rough surfaces, correlated and uncorrelated.
\begin{figure}[!h]
\includegraphics [width=8.0cm] {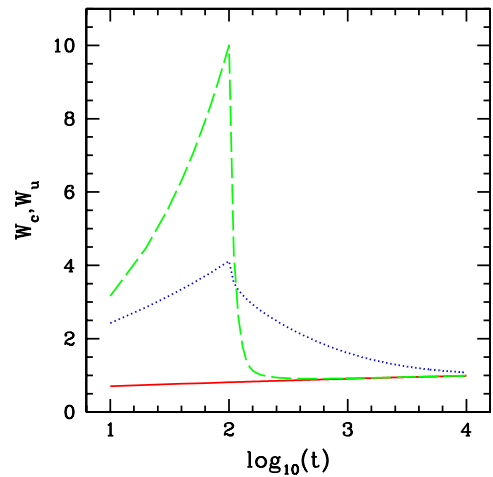}
\caption{(Color online) Roughness evolution in F1 ($W_u$ - red full line), UG-F1 and ETCH-F1
($W_c$ - green dashed and blue dotted lines, respectively) changes at $s=100$.}
\label{fig6}
\end{figure}

Fig. 6 compares the roughness evolution in UG-F1 and ETCH-F1 changes occurring at $s=100$.
In UG-F1, the initial roughness corresponds to the thickness of $10$ layers. It relax to a value
near the unperturbed system after the deposition of $100$ layers or less.
In ETCH-F1, the initial surface is less rough: $W\sim 4$, which corresponds to less than $1nm$ in
a metal or semiconductor surface and possibly some nanometers for larger molecules.
However, the roughness relaxes to a value close to the
unperturbed value only after the deposition of $\sim {10}^4$ layers.
The predicted relaxation exponent for UG-EW is $\gamma =-2$ (second order in $\Delta G\left( k^z s\right)$)
and the relaxation exponent for KPZ-EW is $\gamma =-1$, which suggest a faster decay of
$\Delta W^2$ in the former. An additional reason for the delay in the latter is the larger
crossover time $t_c\sim s^{1.24}\approx 300$.

These results may be very important for the growth of thin films in rough substrates, particularly
when there is some initial pattern or correlated roughness. An investigation of these features in
diffusion dominated growth is certainly desirable.

\section{Conclusion}
\label{conclusion}

The relaxation of the roughness of an interface after a sudden change in the dynamics involving EW
and KPZ growth was studied numerically with lattice models and via scaling arguments. All changes
were considered in the growth regimes of those models, so that power law relaxation is observed in the
square roughness difference $\Delta W^2$ between the changed and the unperturbed systems.

The previous analytical solutions for the EW-EW changes are reviewed and leads to a definition of
a function $\Delta G\left( k^z s\right)$ that contains the basic information on the type of
change of the parameters of the EW equation. Changes in the noise amplitude, with constant interface
tension, give a leading term $\Delta G\left( x \right)\sim x$ (first order), while changes only
in the interface tension give second order dominant term in $\Delta G\left( x \right)$.
The first scaling is also realized when the initial roughness is very small compared to the
unperturbed growth. A hypothesis that $\Delta W^2$ is dominated by a time shift of
the final dynamics is introduced and matches that scaling.

The general form of $\Delta W^2$ in EW-EW changes is extended to KPZ-KPZ changes, which implies
the assumption that $\Delta W^2$ is not dominated by coupling of different modes. The corresponding
function  $\Delta G\left( k^z s\right)$ is also assumed to be of first order.
The predicted relaxation exponents are in good agreement with simulation results in $d=1$ and $d=2$.
Comparison with the recently calculated aging properties of local response functions show
significant differences from the present global response in $d=2$.

KPZ-EW changes are cases in which a faster dynamics is changed to a slower one, corresponding to an
increase in the dynamic exponent. Thus, the time of crossover to the final dynamics is much larger
than the time $s$ of growth with the initial dynamics. We introduce a scaling approach for
the relaxation in those changes, which is also in good agreement with numerical results in $d=1$ and $d=2$.
The hypothesis of $\Delta W^2$ dominated by a time shift of the final dynamics fails in $d=2$
due to the very small EW roughness.

We also compared EW smoothing of initially correlated (KPZ)
and uncorrelated surfaces to illustrate the much slower relaxation in the former.
This may be relevant for thin film growth in rough substrates and may motivate future studies
of the same type of sudden change in growth dominated by surface diffusion.

\acknowledgments
F. Reis acknowledges support from CNPq and FAPERJ (Brazilian agencies).

%~~~~~~~~~~~~~~~~~~~~~~~~~~~~~~~~~~~~~~~~~~~~~~~~~~~~~~~~~~~~~~~~~~~~~~~~~~~
%~~~~~~~~~~~~~~~~~~~ REFERENCES ~~~~~~~~~~~~~~~~~~~~~~~~~~~~~~~~~~~~~~~~~~
%~~~~~~~~~~~~~~~~~~~~~~~~~~~~~~~~~~~~~~~~~~~~~~~~~~~~~~~~~~~~~~~~~~~~~~~~~~~

%~~~~~~~~~~~~~~~~~~~~~~~~~~~~~~~~~~~~~~~~~~~~~~~~~~~~~~~~~~~~~~~~~~~~~~~~~~~
%~~~~~~~~~~~~~~~~~~~ REFERENCES ~~~~~~~~~~~~~~~~~~~~~~~~~~~~~~~~~~~~~~~~~~
%~~~~~~~~~~~~~~~~~~~~~~~~~~~~~~~~~~~~~~~~~~~~~~~~~~~~~~~~~~~~~~~~~~~~~~~~~~~

%\section*{References}
%

\end{document}